\documentclass[10pt,a4paper]{article}
\usepackage[tbtags]{amsmath}
\usepackage[linktocpage]{hyperref}
\hypersetup{colorlinks, linkcolor={blue!60!black}, citecolor={blue!0!black}, urlcolor={blue!80!black}}
\usepackage{amssymb,amsthm}
\usepackage{amsfonts}
\usepackage{dsfont}
\usepackage{graphicx}
\graphicspath{{figs/}}
\usepackage[left=2cm,right=2cm,top=0.5cm,bottom=1.3cm,includeheadfoot]{geometry}
\usepackage{indentfirst}
\usepackage{color}
\usepackage{cite}
\usepackage{mathtools}
\usepackage{tikz-cd}
\usepackage{enumitem}
\usepackage{comment}

\usepackage[normalem]{ulem}

\newcommand{\red}{\textcolor{red}}

\newcommand{\Tr}{\mbox{Tr}}
\newcommand{\be}{\begin{equation}}
\newcommand{\ee}{\end{equation}}
\newcommand{\de}{\mbox{d}}

\newcommand{\pa}{\partial}
\newcommand{\pha}{\phantom{a}}

\numberwithin{equation}{section}

\renewcommand*{\thefootnote}{\fnsymbol{footnote}}
\begin{document}
\begin{flushright} KCL-PH-TH-2025-49\end{flushright}

	\begin{center}
		{\Large\bf A bigravity model from noncommutative geometry
		}
		 \vskip 5mm
		 {\large
		 	Marco de Cesare$^{1,2,}$\footnote{e-mail address: {\tt marco.decesare@na.infn.it}}, Mairi Sakellariadou$^{3,}$\footnote{e-mail address: {\tt mairi.sakellariadou@kcl.ac.uk}} and Araceli Soler Oficial$^{4,}$\footnote{e-mail address: {\tt araceli.soler@ehu.eus}}}
		 \vskip 3mm
		 {\sl $^1$ Scuola Superiore Meridionale, Largo San Marcellino 10, 80138 Napoli, Italy}\\\vskip 1mm
         {$^2$ INFN, Sezione di Napoli, Italy}\\\vskip 1mm
         {\sl $^3$ Theoretical Particle Physics and Cosmology Group, Physics Department, King’s College London, University of London, Strand, London WC2R 2LS, UK}\\\vskip 1mm
         {\sl $^4$ Department of Physics and EHU Quantum Center, University of the Basque Country UPV/EHU,\\
		 	Barrio Sarriena s/n, 48940 Leioa, Spain}
		
	\end{center}
	
\setcounter{footnote}{0}
\renewcommand*{\thefootnote}{\arabic{footnote}}

	\begin{abstract}
Noncommutative gravity, based on a twist-deformation of the differential geometry of spacetime and a first-order formulation of the dynamics, requires additional gravitational degrees of freedom as well as an enlargement of the gauge group of Lorentz transformations of the tetrad frame. As such, it offers a theoretical playground to build fundamentally motivated extensions to general relativity. The dynamical degrees of freedom include a ${\rm GL}(2,\mathbb{C})$ gauge connection and two independent tetrads.  The theory allows for interaction terms between the two tetrads, whose structure displays some similarities with ghost-free bigravity. The extra gravitational degrees of freedom survive in the commutative limit. We show the effective action obtained in this limit, discuss its symmetries, and compare it with other bigravity theories. The dynamics of homogeneous and isotropic cosmological solutions split into two branches. One is characterized by a constant and purely spatial curvature two-form. The other displays a richer gauge freedom, and the Hamiltonian analysis of the dynamics reveals three extra first-class constraints in addition to the generator of time reparametrizations.
	\end{abstract}

\section{Introduction}
The existence of a fundamental length scale implies that spacetime events cannot be localized with infinite accuracy, which results in spacetime uncertainty relations~\cite{Doplicher:1994zv}. Mathematically, this can be modelled by means of a noncommutative deformation of the manifold structure in the framework of twist differential geometry \cite{Aschieri:2005yw,Aschieri:2005zs,Aschieri:2006kc,Aschieri:2007sq,Aschieri_2009}. This allows for the construction of noncommutative theories obtained as deformations of given commutative theories. In the context of gravity, noncommutative geometry offers a natural framework to build fundamentally motivated extensions of Einstein's theory of general relativity. This study aims to explore the connection between noncommutative geometry and bimetric gravity models, which emerge rather naturally in the twist-deformation approach.

In this work, we analyse the cosmological dynamics of a noncommutative gravity model proposed in Ref.~\cite{deCesare:2018cjr} as an extension of \cite{Aschieri:2009ky}, based on a first-order formulation of the dynamics. In this framework, the basic dynamical variables are represented by a ${\rm GL}(2,\mathbb{C})$ gauge connection and a pair of independent tetrad fields. The noncommutative structure is introduced by means of an Abelian twist. Both the symmetry enlargement and the additional degrees of freedom follow from the requirement of mathematical consistency between the twisted differential geometry and gauge symmetry. In this model, there are two distinct types of deviations from general relativity: (i) effects due to the underlying noncommutative structure of spacetime, which become most relevant at small scales; (ii) effects due to the extra dynamical fields, which in principle may play a role also at large scales, where spacetime commutativity is recovered. In Ref.~\cite{Aschieri:2009ky}, this second kind of effects is ruled out at the outset by imposing a `charge conjugation condition'. A different strategy was adopted in Ref.~\cite{Aschieri:2011ng,Aschieri:2012in} to achieve an identical goal, where the Seiberg-Witten map is used to map the noncommutative theory to a commutative one \cite{Seiberg:1999vs}, which results in a higher-derivatives theory with deviations from general relativity that are second-order in the noncommutative deformation parameter. A similar strategy has been adopted in subsequent extensions with ${\rm SO}(2,3)$ symmetry \cite{DimitrijevicCiric:2016qio,Dimitrijevic:2014iwa}.

Here we propose a different approach and retain all of the extra degrees of freedom required by the noncommutative theory, to investigate their role in the dynamics at large scales. We further enrich the dynamics including interaction terms between the two tetrads. The type of interactions that are compatible with the symmetries of the model turn out to be structurally similar to those of ghost-free bigravity \cite{Hassan:2011zd,Hinterbichler:2012cn}. The main difference is that in our case there is a single dynamical gauge connection, whereas in bigravity there are two independent Lorentzian spin connections, one associated with each tetrad. A model closer to ours, with a single Lorentzian spin connection and two independent tetrads, was proposed, in a different context, in Ref.~\cite{Alexandrov:2019dxm}. There, it was shown that, for a large class of backgrounds, the model only allows for a massless graviton as a propagating degree of freedom while the extra degrees of freedom are hidden \cite{Alexandrov:2019dxm}. However, there are some important differences between Ref.~\cite{Alexandrov:2019dxm} and the model at hand (specifically, the values of the independent couplings and the presence of additional dynamical constraints), that prompt an independent analysis.

Within the context of twist differential geometry, the connection with bigravity has been pointed out first in Ref.~\cite{Chamseddine:2003we}; however, the proposed structure of the interaction terms is different from ours and less general. The connection between noncommutative geometry and bimetric gravity has also been explored in the spectral triple approach, based on an almost commutative geometry with two Friedmann-Lema{\^i}tre-Robertson-Walker (FLRW) sheets  \cite{Sitarz:2019trm,Bochniak:2020sbj,Bochniak:2022nxu,Bochniak:2022hhq,Bochniak:2023zqg}, which yields interaction terms with a different structure compared to ghost-free bigravity.

The paper is organized as follows. Section~\ref{Sec:Noncommutative-gravity} reviews the four-dimensional noncommutative gravity theory of Ref.~\cite{deCesare:2018cjr}, constructed via an Abelian twist, and introduces its fundamental structures, namely the bitetrad, the ${\rm GL}(2,\mathbb{C})$ gauge connection, and the deformed wedge product. In Section~\ref{Sec:Commutative-limit}, we consider the commutative limit of the theory and compare the resulting action with two other bigravity models in the literature: the Hassan–Rosen (ghost-free) bimetric gravity theory, and the Alexandrov–Speziale bigravity action with a single spin connection. The corresponding equations of motion are derived in Section~\ref{Sec:field-equations} and subsequently specialized to homogeneous and isotropic cosmologies in Section~\ref{Sec:NGC-cosmo}, where two distinct branches of solutions are identified. In Section~\ref{Sec:SymRed}, we study cosmological dynamics of the symmetry-reduced theory on the physically relevant branch of solutions, and perform the Hamiltonian analysis to examine the gauge symmetries of the system. A comparison between the resulting cosmological dynamics and those of the Alexandrov–Speziale model of Ref.~\cite{Alexandrov:2019dxm} is presented in the Appendix~\ref{Sec:ComparisonAS}. Finally, Section~\ref{Sec:Discussion} provides a concluding discussion of the results obtained throughout the paper.\\

\noindent
{\bf Conventions:} We assume the metric signature $(-+++)$ and work with physical units such that the speed of light and the reduced Planck constant $\hbar$ are both equal to 1.
$\eta_{IJ}$ denotes the Minkowski metric in the tetrad frame. The Levi-Civita symbol with flat indices is normalized such that $\epsilon_{0123}=1$\,, whereas for the antisymmetric tensor density with spacetime indices we take $\tilde{\epsilon}^{\,0123}=1$\,.
We define $\tilde{\epsilon}^{abc}\coloneqq\tilde{\epsilon}^{\,0abc}$\,. The antisymmetrization of indices is denoted by square brackets and includes a factor $1/2$\,, that is, $T_{[ab]}\coloneqq\frac{1}{2}(T_{ab}-T_{ba})$\,.

\section{Noncommutative gravity with bimetric interactions}\label{Sec:Noncommutative-gravity}

An action for noncommutative gravity in four dimensions based on an Abelian twist of the Moyal-Weyl type was formulated in Ref.~\cite{Aschieri:2009ky}, where it was constructed as a deformation of the first-order formulation of general relativity (GR). A generalization of the theory including bimetric interactions and the Holst term was proposed in Ref.~\cite{deCesare:2018cjr}, whose action reads, in the pure gravity case
\be\label{Eq:ActionNCG}
\begin{split}
S_{\rm NCG}[{\bf e},{\boldsymbol \omega}]=\frac{i M_{\rm Pl}^2}{8}\int \Tr\left[{\bf e}\wedge_{\star}{\bf e}\wedge_{\star}\left( *{\bf F}+\frac{1}{\beta}{\bf F}\right) \right]+  i\,c_1 \int\Tr\left[{\bf e}\wedge_{\star}{\bf e}\wedge_{\star}{\bf e}\wedge_{\star}{\bf e}\,\gamma_5\right]\\
+\int c_2 \Tr\left[{\bf e}\wedge_{\star}{\bf e}\wedge_{\star}{\bf e}\wedge_{\star}{\bf e}\right] +c_3 \left(\Tr\left[{\bf e}\wedge_{\star}{\bf e}\right]\right)^2-\frac{c_4}{2^4}\left(\Tr\left[{\bf e}\wedge_{\star}{\bf e}\,\gamma_5\right]\right)^2~.
\end{split}
\ee
Here, $\wedge_{\star}$ denotes the deformed wedge product \cite{Aschieri_2009,Aschieri:2009ky}, whose properties are reviewed below in this section. The $c_n$ are free coupling constants and the dynamical variables are represented by the bitetrad ${\bf e}$ and the ${\rm GL}(2,\mathbb{C})$ gauge connection ${\boldsymbol \omega}$, respectively defined as\footnote{Note the imaginary prefactor of $\tilde{e}^I$ in the definition \eqref{Eq:BitetradDef}, unlike Refs.~\cite{Aschieri:2009ky,deCesare:2018cjr} where the prefactor is real. This redefinition ensures that, in the commutative limit, both $e^I$ and $\tilde{e}^I$ contribute with the same sign to the kinetic term in the action \eqref{Eq:NCG-Action}.}.
\begin{subequations}
\begin{align}
{\bf e} &= e^I \gamma_I+i\,\tilde{e}^I \gamma_I\gamma_5~,\label{Eq:BitetradDef}\\
{\boldsymbol \omega} &=\frac{1}{2}\omega^{IJ}\Gamma_{IJ}+\omega \mathds{1}+\tilde{\omega}\gamma_5~,
\end{align}
\end{subequations}
where $\gamma_I$ are the Dirac matrices spanning the four-dimensional Clifford algebra $\{\gamma_I,\gamma_J\}=2\eta_{IJ}$\,, the chirality matrix is $\gamma_5\coloneqq i\gamma_0\gamma_1\gamma_2\gamma_3$\,, and $\Gamma_{IJ}\coloneqq\frac{i}{4}[\gamma_I,\gamma_J]$ denote the Lorentz algebra generators.
The field strength is defined as 
\be
{\bf F}({\boldsymbol \omega})=\de {\boldsymbol\omega}-i{\boldsymbol \omega}\wedge_{\star}{\boldsymbol \omega}~,
\ee
where ${\rm d}$ denotes the exterior derivative.The Hodge dual of the field strength can be expressed in this notation as $*{\bf F}=-i{\bf F}\gamma_5$\,. Moreover, ${\bf F}$ can be decomposed onto the generators of the Lie algebra of ${\rm GL}(2,\mathbb{C})$ as
\be
{\bf F} =\frac{1}{2}F^{IJ}\Gamma_{IJ}+f\mathds{1}+\tilde{f}\gamma_5~,
\ee
where
\begin{subequations}\label{Eq:field-strength-components}
\begin{align}
F^{IJ} &=\de \omega^{IJ}+\frac{1}{2}\left(\omega^I_{\pha K} \wedge_{\star} \omega^{KJ}-\omega^J_{\pha K} \wedge_{\star} \omega^{KI}\right)-\frac{i}{2}(\omega^{IJ}\wedge_{\star}\omega+\omega\wedge_{\star}\omega^{IJ})+ \frac{1}{2}\left(* \omega^{IJ}\wedge_{\star}\tilde{\omega}+\tilde{\omega}\wedge_{\star}* \omega^{IJ}\right)~,  ~ \\
f &=\de\omega-\frac{i}{8}(\omega^{IJ}\wedge_{\star} \omega_{IJ})-i(\omega\wedge_{\star}\omega+\tilde{\omega}\wedge_{\star}\tilde{\omega})  ~,\\
\tilde{f} &= \de \tilde{\omega}+\frac{1}{8}\omega^{IJ}\wedge_{\star} * \omega^{IJ}-i(\omega\wedge_{\star}\tilde{\omega}+\tilde{\omega}\wedge_{\star}\omega)~.
\end{align}
\end{subequations}
Note that, while $\omega^{IJ}$ corresponds to the standard ${\rm SO}(3,1)$ spin connection, the remaining Abelian components $\omega$, $\tilde{\omega}$ do not have an analogue in GR. Moreover, all of them contribute to the various components of the field strength \eqref{Eq:field-strength-components}, and only decouple in the commutative limit.

The deformed wedge product satisfies the following properties: (i) associativity, i.e.~given three differential forms $\xi$, $\eta$, $\tau$, we have $(\xi\wedge_{\star}\eta)\wedge_{\star}\tau=\xi\wedge_{\star}(\eta\wedge_{\star}\tau)$\,; (ii) graded Leibniz rule, $\de (\sigma \wedge_{\star}\tau)= \de \sigma \wedge_{\star}\tau + (-1)^{\rm deg(\sigma)}\sigma\wedge_{\star}\de\tau$\,; (iii) graded ciclicity, $\int \sigma \wedge_{\star}\tau = (-1)^{\rm deg(\sigma)deg(\tau)}\int \tau \wedge_{\star}\sigma$, up to boundary terms, where ${\rm deg}(\sigma)+{\rm deg}(\tau)=4$\,, the number of spacetime dimensions. These properties ensure that the field equations can be derived by varying the action~\ref{Eq:ActionNCG}, in close analogy with the commutative case.
The following asymptotic expansion  holds for the $\wedge_{\star}$ product of differential forms
\be
\xi \wedge_\star \eta=\xi \wedge \eta+\frac{i}{2}\theta^{\alpha\beta}\mathcal{L}_{X_\alpha}\xi\wedge\mathcal{L}_{X_\beta}\eta+\frac{1}{2!}\left(\frac{i}{2}\right)^2\theta^{\alpha\beta}\theta^{\rho\sigma}\mathcal{L}_{X_\rho}\mathcal{L}_{X_\alpha}\xi\wedge\mathcal{L}_{X_\sigma}\mathcal{L}_{X_\beta}\eta +\mathcal{O}(\theta^3) ~,
\ee
where $\theta^{\alpha\beta}$ is a constant antisymmetric matrix parametrizing the noncommutative deformation.
The vector fields $X_\alpha$ are nondynamical and mutually commuting, $[X_\alpha,X_\beta]=0$\,, i.e.~the twist is Abelian. For further technical details on the mathematical properties of the Abelian twist, the reader is referred to Ref.~\cite{Aschieri:2009ky} and references therein. The physical dimensions of $\theta^{\alpha\beta}$ are ${\rm (length)}^2$, which introduces a new fundamental length scale (which is in principle independent from the Planck length $\ell_{\rm Pl}\sim \sqrt{8\pi \hbar G}$, and may be treated as an additional fundamental constant).

By construction, the action \eqref{Eq:ActionNCG} is invariant under diffeomorphisms and under $\star$-deformed ${\rm GL}(2,\mathbb{C})$ gauge  \cite{Aschieri:2009ky, deCesare:2018cjr}
\begin{subequations}
\begin{align}
 {\bf e}&\rightarrow  {\bf e} +i [ {\bf e},{\boldsymbol \epsilon}]_{\star}~,\label{Eq:transform_bitetrad} \\
 {\boldsymbol \omega}&\rightarrow  {\boldsymbol \omega} -  \left( \de {\boldsymbol \epsilon} - i [ {\boldsymbol \omega},{\boldsymbol \epsilon}]_{\star}\right)~, \\
  {\bf F}&\rightarrow  {\bf F} +i [ {\bf F},{\boldsymbol \epsilon}]_{\star}~.  
 \end{align}
\end{subequations} 
Moreover, the action is also invariant under $\star$-diffeomorphisms, defined as the composition of an ordinary diffeomorphism with the $\star$-deformation. However, it is not background-independent, due to the non-dynamical nature of the vector fields $X_\alpha$ that enter the definition of the twist.

The gauge group ${\rm GL}(2,\mathbb{C})$ is an extension of the Lorentz group ${\rm SO}(3,1)$. In the first-order formulation of GR, the latter is associated with the invariance of the dynamics under local Lorentz transformations of the tetrad frame. Although the matrices $\mathds{1}$ and $\gamma_5$ are not elements of the Lie algebra of ${\rm SO}(3,1)$, they arise through the anticommutators of Lie algebra generators $\Gamma_{IJ}$\,. We note that the $\star$-commutator of infinitesimal gauge transformations can be decomposed as a sum of two terms, one involving the commutator of Lie-algebra generators, one involving their anticommutators. Hence, in order to preserve a notion of gauge invariance, the noncommutative extension of GR necessarily leads to an enlargement of the Lie algebra.

\section{Commutative limit and comparison with bimetric gravity theories}\label{Sec:Commutative-limit}

The noncommutative structure of spacetime becomes most relevant at small distance scales. However, at low energies and at macroscopic length scales, spacetime is well-approximated by a commutative manifold. In this regime, it makes sense to investigate the commutative limit of the noncommutative gravity (NCG) theory \eqref{Eq:ActionNCG}, whereby the deformation parameter tends to zero, $\theta^{\alpha\beta}\to 0$ \cite{deCesare:2018cjr}. Thus, we obtain
\begin{equation}\label{Eq:NCG-Action}
\begin{split}
     &S^{\theta\to0}_{\rm NCG}[e^I,\tilde{e}^I,\omega^{IJ},A]= \frac{M_{\rm Pl}^2}{2}\int \big[P_{IJKL}(e^I\wedge e^J+\tilde{e}^I\wedge \tilde{e}^J)\wedge F^{KL}(\omega)-4 e^{I}\wedge \tilde{e}_{I}\wedge\text{d}A\big]\\
     &+4 c_1\int \epsilon_{IJKL}(e^I\wedge e^J\wedge e^K\wedge e^L+\tilde{e}^I\wedge \tilde{e}^J\wedge \tilde{e}^K\wedge \tilde{e}^L+2e^I\wedge e^J\wedge\tilde{e}^K\wedge \tilde{e}^L) -4 c_4\int e^{I}\wedge\tilde{e}_I\wedge e^J\wedge\tilde{e}_J~,
\end{split}
\end{equation}
where $P_{IJKL}\coloneqq \frac{1}{2}\varepsilon_{IJKL}+\frac{2}{\beta}\eta_{I[K} \eta_{L]J}$ and $A\coloneqq -i\omega+\tilde{\omega}/\beta$ is a one-form associated to a ${\rm U}(1)$ gauge symmetry. We note that the interaction terms with couplings $c_2$, $c_3$ in the action \eqref{Eq:ActionNCG} are exactly vanishing in the commutative limit. In the following, we will refer to the action obtained in the commutative limit \eqref{Eq:NCG-Action} as the NCG action for brevity.

Although the NCG action \eqref{Eq:NCG-Action} is obtained in the commutative limit, it is quite different from the first-order formulation of GR. This is due to the presence of extra dynamical fields, which survive in the limit, that are associated with the enlarged symmetry ${\rm GL}(2,\mathbb{C})$. We stress that the introduction of such extra fields is necessary to ensure consistency between gauge symmetry and the underlying noncommutative structure. Non-trivial interaction terms further enrich the structure of the theory. We observe that deviations from GR persist also in the absence of interaction terms between tetrads (i.e., in the special case $c_1=c_4=0$). Moreover, such deviations are not a peculiarity of the pure gravity theory here considered, and in fact persist also in the presence of matter couplings, unless further conditions are imposed on the dynamical fields. In particular, in Ref.~\cite{Aschieri:2009ky}, where fermionic couplings are included, the continuity of the $\theta\to0$ limit is ensured by imposing a suitable charge conjugation condition on the extra dynamical fields (here denoted $\tilde{e}^I$, $\omega$, $\tilde{\omega}$), such that they vanish in the limit.

The commutative action \eqref{Eq:NCG-Action} retains the ${\rm GL}(2,\mathbb{C})$ gauge symmetry of \eqref{Eq:ActionNCG}. 
An infinitesimal gauge transformation ${\rm GL}(2,\mathbb{C})$ is generated by ${\boldsymbol \epsilon} =\frac{1}{2}\epsilon^{IJ}(x)\Gamma_{IJ}+ \epsilon(x) \mathds{1} + \tilde{\epsilon}(x) \gamma_5$\,. In the commutative limit, the Lorentz subalgebra generated by $\Gamma_{IJ}$\,, and the Abelian subalgebras generated by $\mathds{1}$ and $\gamma_5$ all commute with one another.
The effects of such a gauge transformation on the two independent components of the bitetrad \eqref{Eq:BitetradDef} can be computed using Eq.~\eqref{Eq:transform_bitetrad}in the commutative limit (whereby $[\;\cdot\;,\;\cdot\;]_{\star}\to [\;\cdot\;,\;\cdot\;]$), which gives
\be\label{Eq:gauge-transformation-tetrad-commutativelimit}
\delta e^I =\epsilon^I_{\;J}e^J+2\tilde{\epsilon}\,\tilde{e}^I      ~, \quad
\delta \tilde{e}^I =\epsilon^I_{\pha J}\tilde{e}^J-2\tilde{\epsilon}\,e^I   ~.
\ee
Thus, $\gamma_5$ generates an infinitesimal rotation that mixes $e^I$ and $\tilde{e}^I$. This is in addition to the infinitesimal Lorentz transformation parametrized by $\epsilon^I_{\;J}$\,. On the other hand, $\mathds{1}$ has no effect on the bitetrad.

We observe that, upon defining $E^I=e^I+i\tilde{e}^I$ and its complex conjugate
$\overline{E}^I=e^I-i\tilde{e}^I$, the action \eqref{Eq:NCG-Action} can be rewritten in a more compact form
\begin{equation}\label{Eq:NCG-Action-compact}
\begin{split}
     S^{\theta\to0}_{\rm NCG}= &\frac{M_{\rm Pl}^2}{2}\int \left[\frac{1}{2}P_{IJKL}(E^I\wedge \overline{E}^J+\overline{E}^I\wedge E^J)\wedge F^{KL}(\omega)-2i E^{I}\wedge \overline{E}_{I}\wedge\text{d}A\right]\\
     &+4 c_1\int \epsilon_{IJKL}E^I\wedge E^J\wedge \overline{E}^K\wedge \overline{E}^L +c_4\int \left(E^{I}\wedge\overline{E}_I\right)^2~.
\end{split}
\end{equation}
Written in this form, the action is manifestly invariant under $U(1)$ gauge transformations $E^I\to e^{-i\varphi}E^I$\,, $\overline{E}^I\to e^{i\varphi}\overline{E}^I$. This is equivalent to the symmetry that mixes $e^I$ and $\tilde{e}^I$ which is included in \eqref{Eq:gauge-transformation-tetrad-commutativelimit}.

The structure of the action \eqref{Eq:NCG-Action} displays several similarities with the action of the ghost-free bimetric theory formulated by Hassan and Rosen \cite{Hassan:2011vm, Hassan:2011tf}, which modifies general relativity by considering two non-linearly interacting spin-2 fields. In its tetrad formulation, the action of the Hassan-Rosen theory reads \cite{Hinterbichler:2012cn}
\begin{equation}\label{Eq:Bimetric-Action}
\begin{split}
     &S_{\rm HR}[e^I,\tilde{e}^I,\omega^{IJ},\tilde{\omega}^{IJ}] = \frac{M_g^2}{4}\int \epsilon_{IJKL}e^I\wedge e^J\wedge F^{KL}(\omega)+\frac{M_f^2}{4}\int \epsilon_{IJKL}\tilde{e}^I\wedge \tilde{e}^J\wedge F^{KL}(\tilde{\omega})\\
     &\qquad -M_g^2\int \epsilon_{IJKL}\bigg(\frac{\beta_4}{4!}\tilde{e}^I\wedge \tilde{e}^J\wedge \tilde{e}^K\wedge \tilde{e}^L +\frac{\beta_3}{3!}e^I\wedge \tilde{e}^J\wedge \tilde{e}^K\wedge \tilde{e}^L+\frac{\beta_2}{(2!)^2}e^I\wedge e^J\wedge \tilde{e}^K\wedge \tilde{e}^L\\
     &\qquad +\frac{\beta_1}{3!}e^I\wedge e^J\wedge e^K\wedge \tilde{e}^L+\frac{\beta_0}{4!}e^I\wedge e^J\wedge e^K\wedge e^L\bigg)~,
\end{split}
\end{equation}
where the $\beta_n$ are arbitrary constants with physical dimensions of ${\rm(mass)}^2$ and $M_g$ and $M_f$ are the Planck masses associated to $e^I$ and $\tilde{e}^I$, respectively.
The main difference between \eqref{Eq:NCG-Action} and \eqref{Eq:Bimetric-Action} is that in the Hassan-Rosen theory there exist two distinct spin connections, which are dynamically independent, leading to a kinetic term which comprises two independent copies of the Einstein-Hilbert term of GR (in its first-order form). Then, the two tetrads couple to one another through the interaction terms, which are built in such a way that the Boulware-Deser ghost is absent. In the NCG model \eqref{Eq:NCG-Action}, while the interaction terms retain a structure that is similar to the Hassan-Rosen theory, two additional terms are introduced, describing interactions of a new kind. Finally, \eqref{Eq:NCG-Action} differs from \eqref{Eq:Bimetric-Action} in the inclusion of the Holst term, which becomes important in the presence of torsion, which is sourced e.g.~by fermions.

Another related theory has been proposed by Alexandrov and Speziale in Ref.~\cite{Alexandrov:2019dxm}, which is constructed in analogy with the Hassan-Rosen bimetric action \eqref{Eq:Bimetric-Action} although with a crucial difference: there are two tetrads but a single spin connection associated with local Lorentz symmetry. Explicitly, the theory is described by the following action\footnote{In order to match the notation in \cite{Hinterbichler:2012cn} we have performed the change $\beta_n\to\beta_{4-n}$ with respect to the Alexandrov-Speziale action definition.}
\begin{equation}\label{Eq:Speziale-Action}
\begin{split}
     &S_{\rm AS}[e^I,\tilde{e}^I,\omega^{IJ}]=  \frac{M_{\rm Pl}^2}{4}\int \epsilon_{IJKL}(e^I\wedge e^J+\tilde{e}^I\wedge \tilde{e}^J)\wedge F^{KL}(\omega) -M_{\rm Pl}^2\int \epsilon_{IJKL}\bigg(\frac{\beta_4}{4!}\tilde{e}^I\wedge \tilde{e}^J\wedge \tilde{e}^K\wedge \tilde{e}^L\\
     & +\frac{\beta_3}{3!}e^I\wedge \tilde{e}^J\wedge \tilde{e}^K\wedge \tilde{e}^L+\frac{\beta_2}{(2!)^2}e^I\wedge e^J\wedge \tilde{e}^K\wedge \tilde{e}^L+\frac{\beta_1}{3!}e^I\wedge e^J\wedge e^K\wedge \tilde{e}^L +\frac{\beta_0}{4!}e^I\wedge e^J\wedge e^K\wedge e^L\bigg)~.
\end{split}
\end{equation}
We note that the interaction terms in \eqref{Eq:NCG-Action} correspond to setting $\beta_1=\beta_3=0$\,, $\beta_0=\beta_4=3\beta_2$ in the action \eqref{Eq:Speziale-Action}. Moreover, the term involving the one-form $A_{\mu}$ as well as the interaction parametrized by $c_4$ do not have an analog in Eq.~\eqref{Eq:Speziale-Action}.
Nonetheless, there are some similarities between the two theories, \eqref{Eq:NCG-Action} and \eqref{Eq:Speziale-Action}, at the dynamical level, which are discussed in Appendix~\ref{Sec:ComparisonAS}.

\section{Field equations}\label{Sec:field-equations}
In this section, we derive the full set of field equations for the commutative limit of the theory. The variation of the action \eqref{Eq:NCG-Action} with respect to $\omega^{IJ}$ gives
\begin{equation}\label{Eq:Eom-a}
        \mathcal{D}_\omega\left(e^I\wedge e^J+\tilde{e}^I\wedge\tilde{e}^J\right)=0~.
\end{equation}
Here $ \mathcal{D}_\omega$ denotes the exterior covariant derivative,
\begin{equation}
    \mathcal{D}_\omega\left(e^I\wedge e^J+\tilde{e}^I\wedge\tilde{e}^J\right)={\rm d}\left(e^I\wedge e^J+\tilde{e}^I\wedge\tilde{e}^J\right)+\omega^I{}_K\wedge\left( e^K\wedge e^J+ \tilde{e}^K\wedge \tilde{e}^J\right)+\omega^J{}_K\wedge \left(e^I\wedge e^K +\tilde{e}^I\wedge \tilde{e}^K\right)~.
\end{equation}
In analogy with the standard first-order formulation of GR, we will refer to \eqref{Eq:Eom-a} as the \textit{connection equation}. We observe that an identical equation follows from the Alexandrov-Speziale action \eqref{Eq:Speziale-Action} \cite{Alexandrov:2019dxm}. 
The general solution to this equation is quite involved and has been obtained in Ref.~\cite{Beke:2011mu}. 
For an analysis of different families of solutions to the connection equation see \cite{deCesare:2018cjr}.

The variation with respect to the one-form $A$ gives
\begin{equation}\label{Eq:Eom-b}
     \text{d}\left(e^I\wedge\tilde{e}_I\right)=0~.
\end{equation}
One special solution to this equation is $e^I\wedge \tilde{e}_I=0$\,, which we will refer to as the  \textit{symmetric vielbeins condition}. In Hassan-Rosen bimetric gravity \cite{
Hinterbichler:2012cn}, this condition follows from a secondary constraint \cite{Alexandrov:2013rxa} and ensures that there exists an equivalence between metric and tetrad formulations.\footnote{Even more, solutions where the symmetric vielbein condition is broken exhibit a ghost \cite{deRham:2015cha}.} In our case, the symmetric vielbeins condition is just a particular solution to \eqref{Eq:Eom-b} and, in general, the theory does not admit a metric formulation. This is in analogy with the model of Ref.~\cite{Alexandrov:2019dxm}. 

Finally, varying the action \eqref{Eq:NCG-Action} with respect to $e^I$ and $\tilde{e}^I$ one obtains, respectively
\begin{subequations}
\begin{align}
     &M_{\rm Pl}^2\left(P_{IJKL}\, e^J\wedge F^{KL}-2\tilde{e}_I\wedge \text{d}A\right)
   +16\, c_1\epsilon_{IJKL}\left(e^J\wedge e^K\wedge e^L+e^J\wedge\tilde{e}^K\wedge\tilde{e}^L\right)-8\,c_4\,\tilde{e}_I\wedge e^J\wedge \tilde{e}_J=0~,\label{Eq:Eom-c} \\
&M_{\rm Pl}^2\left(P_{IJKL}\, \tilde{e}^J\wedge F^{KL}+2 e_I \wedge\text{d}A\right) +16\, c_1\epsilon_{IJKL}\left(\tilde{e}^J\wedge \tilde{e}^K\wedge \tilde{e}^L+e^J\wedge e^K\wedge\tilde{e}^L\right)+8\,c_4\,e_I\wedge e^J\wedge \tilde{e}_J=0~.\label{Eq:Eom-d}
\end{align}
\end{subequations}
It is important to note that when the symmetric vielbeins condition is satisfied, the final term in both equations \eqref{Eq:Eom-c} and \eqref{Eq:Eom-d} vanishes.

As discussed in Ref.~\cite{Alexandrov:2019dxm}, in the absence of interaction terms, the action \eqref{Eq:Speziale-Action} (and thus also \eqref{Eq:NCG-Action}), apart from being invariant under local Lorentz transformations and diffeomorphisms, has two additional gauge symmetries, $\mathcal{S}_1$ and $\mathcal{S}_2$, that at the infinitesimal level act as
\begin{subequations}
    \begin{align}
        & \mathcal{S}_1: \quad e^I_{\mu} \to  e^I_{\mu}+\varepsilon \tilde{e}^I_{\mu}~,\hspace{3.45cm} \tilde{e}^I_{\mu} \to\tilde{e}^I_{\mu}-\varepsilon e^I_{\mu}~,\label{Eq:S1}\\
         & \mathcal{S}_2: \quad e^I_{\mu} \to  e^I_{\mu}+\varepsilon \tilde{e}\left[\frac{1}{2}e^J_{\nu}\tilde{e}^\nu_Je^I_{\mu}-e^I_{\nu}\tilde{e}^\nu_Je^J_\mu\right]~,\quad \tilde{e}^I_{\mu} \to\tilde{e}^I_{\mu}-\varepsilon e\left[\frac{1}{2}\tilde{e}^J_{\nu}e^\nu_J\tilde{e}^I_{\mu}-\tilde{e}^I_{\nu}e^\nu_J\tilde{e}^J_\mu\right]~,\label{Eq:S2}
    \end{align}
\end{subequations}
where $\varepsilon$ is a transformation parameter. In general, these are not symmetries of the full theory \eqref{Eq:Speziale-Action} when interactions are included. However, it turns out that for the action \eqref{Eq:NCG-Action} the symmetry ${\cal S}_1$  in \eqref{Eq:S1} is a gauge symmetry of the full theory and corresponds to the Abelian symmetry generated by $\gamma_5$\,, already noticed in Eq.~\eqref{Eq:gauge-transformation-tetrad-commutativelimit}. 
This is unlike the action \eqref{Eq:Speziale-Action}, where the ${\cal S}_1$ symmetry is broken by interactions. The enhanced symmetry of \eqref{Eq:NCG-Action} compared to \eqref{Eq:Speziale-Action} is due to the particular structure of the bimetric-type interaction term in \eqref{Eq:NCG-Action}, which corresponds to a particular tuning of the $\beta_n$ parameters in \eqref{Eq:Speziale-Action} (in fact, ${\cal S}_1$ would be broken if the relative coefficients between the three terms were different). The enhanced gauge symmetry has non-trivial consequences for the dynamics, as shown in the following sections where we analyze the dynamics of the cosmological sector.

\section{Cosmological dynamics and Hamiltonian analysis}\label{Sec:NGC-cosmo}

In this section, we analyse the dynamics of a cosmological spacetime in the theory at hand.
Let us start by considering two independent metric tensors, $g_{\mu\nu}$ and $f_{\mu\nu}$\,, which are assumed to describe perfectly homogeneous and isotropic cosmologies with consistent time foliations (i.e., the corresponding spatial leaves are both assumed to have normal one-form $\propto {\rm d}t$\,, for a given time function $t$). We also assume spatial flatness for simplicity.
The most general ansatz compatible with these assumptions reads
\begin{equation}\label{Eq:BimetricAnsatz}
    \text{d}s^2_{(g)}=a^2(t)\left(-n^2(t)\text{d}t^2+\text{d}\Vec{x}{\,}^2\right)~,\qquad\text{d}s^2_{(f)}=b^2(t)\left(-c^2(t)n^2(t)\text{d}t^2+\text{d}\Vec{x}{\,}^2\right)~,
\end{equation}
where $a(t)$ and $b(t)$ are the scale factors, $n(t)$ is the lapse for the metric $g_{\mu\nu}$\,, and $c(t)$ parametrizes the relative lapse between the conformal times of the two metrics. Coordinates are denoted as $x^0=t\,,~x^1=x\,,~x^2=y\,,~x^3=z$\,.
The lapse function $n(t)$ is kept completely general, which is necessary to ensure consistency of the Hamiltonian analysis performed in the following. A choice of tetrads corresponding to the metric ansatz \eqref{Eq:BimetricAnsatz} is the following
\begin{equation}\label{Eq:FLRW-ansatz}
    e^I_{\;\mu}=a(t)\left(n(t)\delta^I_{\;0}\delta^0_{\;\mu}+\delta^I_{\;i}\delta^i_{\;\mu}\right),\qquad \tilde{e}^I_{\;\mu}=b(t)\left(n(t)c(t)\delta^I_{\;0}\delta^0_{\;\mu}+\delta^I_{\;i}\delta^i_{\;\mu}\right).
\end{equation}
With this ansatz, the symmetricity condition $e^I\wedge \tilde{e}_I=0$ is automatically satisfied.

The ansatz \eqref{Eq:FLRW-ansatz} is motivated by simplicity. It is possible to consider a more general ansatz for the tetrads, still compatible with \eqref{Eq:BimetricAnsatz}, such that $e^I\wedge \tilde{e}_I\neq0$ and \eqref{Eq:Eom-b} thus gives a non-trivial constraint.
For instance, take
\begin{equation}\label{Eq:FLRW-ansatz-general}
    e^I_{\;\mu}=a(t)\Lambda^I_{\;J}(t)\left(n(t)\delta^J_{\;0}\delta^0_{\;\mu}+\delta^J_{\;i}\delta^i_{\;\mu}\right),\qquad \tilde{e}^I_{\;\mu}=b(t)\left(n(t)c(t)\delta^I_{\;0}\delta^0_{\;\mu}+\delta^I_{\;i}\delta^i_{\;\mu}\right),
\end{equation}
where $\Lambda^{I}_{\;J}(t)$ is a general (time-dependent) Lorentz transformation, parametrizing the relative rotation or boost between the two tetrads. As an example, consider $\Lambda^{I}_{\;J}(t)$ being such that the tetrad $e^I$ is rotated by an angle $\vartheta(t)$ about the $x$-axis with respect to the tetrad $\tilde{e}^I$. In this case, Eq.~\eqref{Eq:Eom-b} gives $\partial_t\left(a(t) b(t) \sin{\vartheta(t)}\right)=0$\,, which imposes a non-trivial restriction on the time-dependence of the rotation angle and makes the analysis of the dynamics significantly more involved. For this reason, in the following we will consider for simplicity the ansatz \eqref{Eq:FLRW-ansatz}, which is a particular case of \eqref{Eq:FLRW-ansatz-general} with $\Lambda^I_{\;J}=\delta^I{}_J$\,, so that the symmetricity condition holds automatically at all times. As a consequence, the remaining equations of motion \eqref{Eq:Eom-a}--\eqref{Eq:Eom-d} take a simpler form too.

Lastly, using the cosmological ansatz \eqref{Eq:FLRW-ansatz} we obtain (omitting the time-dependence to make the notation lighter)
\be\label{Eq:area-2form}
e^I\wedge e^J+\tilde{e}^I\wedge\tilde{e}^J=2n\left(a^2+b^2c\right)\delta^{[I}_{0}\delta^{J]}_i\de t\wedge \de x^i+\left(a^2+b^2\right)\delta^{[I}_{i}\delta^{J]}_j \de x^i\wedge \de x^j~.
\ee
Taking the exterior covariant derivative of this expression and substituting it into the connection equation \eqref{Eq:Eom-a} gives
\begin{equation}\label{Eq:connection-eq-cosmological}
\begin{split}
    \Big[(a\dot{a}+b\dot{b})\delta^{[I}_i\delta^{J]}_j-n(a^2+b^2c)\left(\omega^{[I}_{i}{}_0\,\delta^{J]}_j+\omega^{[I}_{i}{}_j\,\delta^{J]}_0\right)  +(a^2+b^2) & \omega^{[I}_{0}{}_i\,\delta^{J]}_j\Big]  \de t\wedge \de x^i\wedge\de x^j\\
    & +(a^2+b^2)\omega^{[I}_{l}{}_i\,\delta^{J]}_j\de x^l\wedge \de x^i\wedge\de x^j=0~.
\end{split}
\end{equation}
Interestingly, solutions to the connection equation \red{\eqref{Eq:connection-eq-cosmological}} fall into two different branches, depending on whether the two-form \eqref{Eq:area-2form} is purely spatial or has non-zero time-space components (equivalently, whether $a^2+ b^2c$ vanishes or not). The two branches are analyzed separately in the following sections.

\subsection{Branch I}\label{Subsec:BranchII-NCG}

On this branch $a^2+ b^2c\neq 0$\,, and then, from \eqref{Eq:connection-eq-cosmological}, and using the antisymmetry of $\omega^{IJ}$ , the solution to the connection equation \eqref{Eq:Eom-a} reads
\begin{equation}
   \omega_{\mu}^{IJ}=
         h(t) \left(\delta^I_{\;0}\delta^J_{\;\mu}-\delta^I_{\;\mu}\delta^J_{\;0}\right) , 
\end{equation}
where we used the notation
\begin{equation}\label{Eq:Def-h}
     h(t)\coloneqq \frac{a(t)\dot{a}(t)+b(t)\dot{b}(t)}{n(t)\left[a(t)^2+b(t)^2c(t)\right]}~,
    \end{equation}
and a dot represents the derivative with respect to $t$\,. On the other hand, from the field equations \eqref{Eq:Eom-c} and \eqref{Eq:Eom-d} we obtain
\begin{subequations}\label{Eq:Cosmological-eom}
  \begin{equation}\label{Eq:Cosmological-eom-a}
     h^2=-\frac{\tilde{c}_1}{M_{\rm Pl}^2}(a^2+b^2)~,
     \end{equation}
     \begin{equation}\label{Eq:Cosmological-eom-b}
     M_{\rm Pl}^2(2\dot{h}+n\,h^2)+\tilde{c}_1 n\left[3a^2+2b^2(1+2c)\right]=0~,
    \end{equation}
     \begin{equation}\label{Eq:Cosmological-eom-c}
    M_{\rm Pl}^2(2\dot{h}+n\,c\,h^2)+\tilde{c}_1 n\left[3cb^2+a^2(2+c)\right]=0~,
    \end{equation}
     \begin{equation}\label{Eq:Cosmological-eom-d}
    \partial_{[\mu}A_{\nu]}=0~,
    \end{equation}  
\end{subequations}
with $\tilde{c}_1\coloneqq2^5c_1$\,. It follows from Eq.~\eqref{Eq:Cosmological-eom-d} that the field strength associated with the ${\rm U}(1)$ gauge connection $A_\mu$ is identically zero, and thus $A_\mu$ is in a pure gauge configuration (that is, $A_\mu=\pa_\mu \lambda$ for some arbitrary scalar $\lambda$). Moreover, eliminating $\dot{h}$ from \eqref{Eq:Cosmological-eom-b} and \eqref{Eq:Cosmological-eom-c} we obtain \eqref{Eq:Cosmological-eom-a}. This shows that the above equations are not independent, and therefore we can omit e.g.~\eqref{Eq:Cosmological-eom-c} in the following. By differentiating \eqref{Eq:Cosmological-eom-a} and recalling the definition of $h$ in \eqref{Eq:Def-h}, we find that Eq.~\eqref{Eq:Cosmological-eom-b} is also redundant. Hence, treating $h$ as an independent field, the previous system reduces to a set of just two independent equations
\begin{subequations}\label{Eq:Cosmological-eom-final}
\begin{align}
     h  &=\frac{a\dot{a}+b\dot{b}}{n(a^2+b^2c)}~,\label{Eq:Cosmological-eom-final-a}\\
     h^2  &=-\frac{\tilde{c}_1}{M_{\rm Pl}^2}(a^2+b^2)~.\label{Eq:Cosmological-eom-final-b}
\end{align}
\end{subequations}
Given that $h$ is real, from Eq.~\eqref{Eq:Cosmological-eom-final-b} we conclude that, on this branch $c_1<0$ is necessary to ensure consistency of the dynamics.
The general solution to the system \eqref{Eq:Cosmological-eom-final} can be given in terms of three arbitrary functions of time, namely $u(t)$, $v(t)$, and $w(t)$\,: 
\begin{equation}\label{Eq:Solution-NCG-branchI}
   h=u~, \qquad a=\frac{|u|}{\mathcal{C}_1}\cos{v}~,\qquad b=\frac{|u|}{\mathcal{C}_1}\sin{v}~,\qquad c=\frac{|\dot{u}||u|}{w\,u^3\sin^2{v}}-\cot^2{v}~, \qquad n=w~,
\end{equation}
with $\mathcal{C}_1 \coloneqq \sqrt{-\tilde{c}_1}/M_{\rm Pl}$\,. The fact that the general solution is completely expressed in terms of free functions is a peculiar property of the system, which depends on the high degree of gauge symmetries. A detailed comparison with the dynamics in the Alexandrov-Speziale is carried out in the Appendix~\ref{Sec:ComparisonAS}. In Section~\ref{Sec:SymRed}, we further investigate the dynamical structure of the theory in its symmetry-reduced cosmological sector, to shed light on the gauge redundancies displayed by the exact solution \eqref{Eq:Solution-NCG-branchI}.

\subsection{Branch II}\label{Subsec:BranchI-NCG}
Solutions on this branch are such that $a^2+ b^2c=0$\,, whereby the two-form \eqref{Eq:area-2form} does not have a temporal component.
Solving the connection equation \eqref{Eq:connection-eq-cosmological}  with $a^2+ b^2c=0$ gives
\begin{equation}\label{Eq:Connection-branchI-sol}
        a\dot{a}+b\dot{b}  =0~, \qquad \omega^i_{jl}=\omega^I_{0i}  =0~, \qquad \omega^0_{ij}  =\omega^0_{ji}~.
    \end{equation}
Thus, we have six unknown independent components of the spin connection $\omega^I_{\mu J}$\,, namely $\omega^0_{11}$\,, $\omega^0_{12}$\,, $\omega^0_{13}$\,, $\omega^0_{23}$\,, $\omega^0_{22}$\,, $\omega^0_{33}$ that are left to determine. Moreover, from the first equation, we obtain the algebraic constraint $a^2+b^2=\mathcal{C}^2$, with $\mathcal{C}$ an arbitrary real constant. Hence, the evolution of the two scale factors corresponds to the motion of a point particle on a circle with constant radius, with arbitrary velocity. In turn, this constraint implies that the two-form \eqref{Eq:area-2form}, besides being purely spatial on this branch, is also constant. The evolution of the scale factors and the relative lapse can be expressed as, $a=\mathcal{C}\cos\theta(t)$\,, $b=\mathcal{C}\sin\theta(t)$\,, $c=-(\tan\theta(t))^{-2}$\,, with $\theta(t)$ an arbitrary function of time. This arbitrariness in the scale factors is not fixed by the remaining equations of motion, as shown below.

Note that, unlike the first-order formulation of GR, in this case the connection equation \eqref{Eq:Eom-a} does not completely determine the spin connection in terms of the tetrad, but rather imposes restrictions on the tetrads themselves. Hence, to obtain solutions for all the components of the spin connection, one needs to solve the remaining field equations, \eqref{Eq:Eom-c} and \eqref{Eq:Eom-d}, obtained by varying the action \eqref{Eq:ActionNCG} with respect to the two tetrads. Equations \eqref{Eq:Eom-c} and \eqref{Eq:Eom-d} read, after some algebraic manipulations,
\begin{subequations}\label{Eq:Eom-branchI-NCG}
  \begin{equation}\label{Eq:Eom-branchI-NCG-a}
  \omega^0_{11}\omega^0_{22}+\omega^0_{11}\omega^0_{33}+\omega^0_{22}\omega^0_{33}-(\omega^0_{12})^2-(\omega^0_{13})^2-(\omega^0_{23})^2+\frac{3 \tilde{c}_1\mathcal{C}^2}{M_{\rm Pl}^2} =0~,
\end{equation} 
\begin{equation}\label{Eq:Eom-branchI-NCG-b}
    \partial_{[\mu}A_{\nu]}=0~,
\end{equation}
\begin{equation}\label{Eq:Eom-branchI-NCG-c}
    \omega^0_{ii}\omega^0_{jj}-(\omega^0_{ij})^2+\frac{1}{n}\left(\Dot{\omega}^0_{ii}+\Dot{\omega}^0_{jj}\right)+\frac{\tilde{c}_1\mathcal{C}^2}{M_{\rm Pl}^2}=0~,
\end{equation}
\begin{equation}\label{Eq:Eom-branchI-NCG-d}
    \omega^0_{ii}\omega^0_{jk}-\omega^0_{ij}\omega^0_{ik}+\frac{1}{n}\Dot{\omega}^0_{jk}=0~,
\end{equation}
\begin{equation}\label{Eq:Eom-branchI-NCG-e}
    \omega^0_{ii}\omega^0_{jj}-(\omega^0_{ij})^2-\frac{b^2}{a^2 n}\left(\Dot{\omega}^0_{ii}+\Dot{\omega}^0_{jj}\right)+\frac{\tilde{c}_1\mathcal{C}^2}{M_{\rm Pl}^2}=0~,
\end{equation}
\begin{equation}\label{Eq:Eom-branchI-NCG-f}
    \omega^0_{ii}\omega^0_{jk}-\omega^0_{ij}\omega^0_{ik}-\frac{b^2}{a^2n}\Dot{\omega}^0_{jk}=0~.
\end{equation}
\end{subequations}
Here, Latin indices $i,j,k$ are used to represent a cyclic permutation of $1,2,3$\,. Repeated indices are not summed over.
First, from \eqref{Eq:Eom-branchI-NCG-b} we conclude that  $A_\mu$ is pure gauge also on this branch.  Regarding the ${\rm SO}(3,1)$ spin connection, combining \eqref{Eq:Eom-branchI-NCG-a}, \eqref{Eq:Eom-branchI-NCG-c} and \eqref{Eq:Eom-branchI-NCG-e} on one side and \eqref{Eq:Eom-branchI-NCG-d} and \eqref{Eq:Eom-branchI-NCG-f} on the other, we obtain that all the components must be constant. Explicitly, we find
\begin{equation}\label{Eq:sol-omega-branchI}
    \omega^0_{11}=\omega^0_{22}=\omega^0_{33}= \sqrt{\mathcal{C}_1^2\mathcal{C}^2}~, \qquad \omega^0_{12}=\omega^0_{13}=\omega^0_{23}=0~.
\end{equation}
Therefore, the spin connection is real if and only if $c_1<0$\,.

At this stage, we can compute the field strength corresponding to the solution \eqref{Eq:sol-omega-branchI}, obtaining
\begin{equation}\label{Eq:BranchI_FieldStrength}
  F^{IJ}=\mathcal{C}_1^2\mathcal{C}^2\delta^I_{[i}\delta^J_{\;j]} \text{d}x^i\wedge\text{d}x^j~,
\end{equation}
which is a purely spatial and constant 2-form. Given that there are two independent tetrads, but only one spin connection, two possible notions of the torsion tensor are given. The torsion two-form, defined with respect to the tetrad $e^I$ is $T^I\coloneqq \text{d}e^I+\omega^I{}_J\wedge e^J=\left[\sqrt{\mathcal{C}_1^2\mathcal{C}^2}\,n\,a-\dot{a}\right]\delta^I_{\;i}\text{d}t\wedge \text{d}x^i$, whereas using the second tetrad $\tilde{e}^I$ one has an alternative (in principle inequivalent) definition $\tilde{T}^I\coloneqq \text{d}\tilde{e}^I+\omega^I{}_J\wedge \tilde{e}^J=\left[\sqrt{\mathcal{C}_1^2\mathcal{C}^2}\,n\,a^2/b-\dot{b}\right]\delta^I_{\;i}\text{d}t\wedge \text{d}x^i$\,. Let us then decompose the spin connection, following similar steps as in Ref.~\cite{Magueijo:2012ug}: $\omega^{IJ}=\bar{\omega}^{IJ}+C^{IJ}$, where $\bar{\omega}^{IJ}$ is torsion-free with respect to both tetrads (i.e.,~$\text{d}e^I+\bar{\omega}^I{}_J\wedge e^J=\text{d}\tilde{e}^I+\bar{\omega}^I{}_J\wedge\tilde{e}^J=0$), and $C^{IJ}$ is the contorsion one-form that satisfies $T^{I}=C^{IJ}\wedge e_J$ and $\tilde{T}^{I}=C^{IJ}\wedge\tilde{e}_J$ (given that, on this branch, $\dot{a}/a=\dot{b}/bc$):
\begin{equation}\label{Eq:ConnectionSol_Branch1}
    \bar{\omega}^{IJ}=\frac{2\dot{a}}{n\,a}\delta^{[I}_{\,\,0}\delta^{J]}_{\;i} \text{d}x^i~,\qquad C^{IJ}=2\left(\sqrt{\mathcal{C}_1^2\mathcal{C}^2}-\frac{\dot{a}}{n\,a}\right)\delta^{[I}_{\,\,0}\delta^{J]}_{\;i} \text{d}x^i~.
\end{equation}
We stress that $\bar{\omega}^{IJ}$ is compatible with both tetrads, since $\dot{a}/a=\dot{b}/bc$\,. Hence, the spin connection $\omega^{IJ}$ naturally decomposes into a torsion-free part $\bar{\omega}^{IJ}$, which corresponds to the connection compatible with both tetrads, and a contorsion part $C^{IJ}$, which encodes the deviation from metric compatibility.

The result obtained for the field strenght~\eqref{Eq:BranchI_FieldStrength}, should be contrasted with the corresponding solution in GR for an FLRW spacetime,
\begin{equation}\label{Eq:GR_FieldStrength}
    F^{IJ}=2\delta^{[I}_0\delta^{J]}_i\left(\frac{\Ddot{a}\,a\,n-\dot{a}\,a\,\dot{n}-\dot{a}^2n}{a^2n^2}\right)\de t\wedge\de x^i+\left(\frac{\dot{a}}{a\,n}\right)^2\delta^I_i\delta^J_j\de x^i\wedge\de x^j~.
\end{equation}
We observe that the $\de t\wedge\de x^i$ component of \eqref{Eq:GR_FieldStrength} is zero if and only if the Hubble rate $H=\dot{a}/an$ is constant. 
Only in this case, an analogy may be drawn between \eqref{Eq:BranchI_FieldStrength} and \eqref{Eq:GR_FieldStrength}. Nonetheless, in the NCG model the Hubble rate need not be constant and in fact remains completely arbitrary. Only if the Hubble rate equals $ \sqrt{\mathcal{C}_1^2\mathcal{C}^2}$\,, torsion vanishes according to \eqref{Eq:ConnectionSol_Branch1}, and one obtains a de Sitter solution. Hence, given the limited physical relevance of this branch, we will not analyze it further in this work.

\section{Symmetry reduction and canonical analysis}\label{Sec:SymRed}
In this Section, we provide an in-depth analysis of the dynamics of the cosmological background on Branch~I.  First, in Section~\ref{Sec:SymRed_sub1}, we establish a precise correspondence between the full NCG action \eqref{Eq:NCG-Action} and the symmetry-reduced theory, showing that, when the theory at hand is restricted to Branch~I, the {\it principle of symmetric criticality} holds for cosmological spacetimes. That is, the derivation of the equations of motion commutes with symmetry reduction. Note that this is not guaranteed a priori, since the hypotheses of Ref.~\cite{Palais1979} are not satisfied. Next, in Section~\ref{Sec:CanonicalAnalysis}, taking the symmetry-reduced action as a starting point, we build a Hamiltonian description of the system using the Dirac-Bergmann algorithm for constrained systems \cite{dirac1964lectures}. We classify constraints, identify the generators of gauge transformations, and study their action on physical states.

\subsection{Symmetry-reduced action}\label{Sec:SymRed_sub1}

Let us examine whether the {\it principle of symmetric criticality} holds in the NCG model for homogeneous and isotropic spacetimes on Branch I. That is, we investigate whether the dynamics obtained from the restriction of the field equations \eqref{Eq:Eom-a}--\eqref{Eq:Eom-d} to the cosmological ansatz \eqref{Eq:FLRW-ansatz} are equivalent to the Euler-Lagrange equations derived from the symmetry-reduced action. Specifically, we substitute the following ansatz into the action \eqref{Eq:NCG-Action}
\begin{equation}
e^I_{\mu}=a(t)\left[n(t)\delta^I_{\;0}\delta^0_{\;\mu}+\delta^I_{\;i}\delta^i_{\;\mu}\right],\quad \tilde{e}^I_{\mu}=b(t)\left[n(t)c(t)\delta^I_{\;0}\delta^0_{\;\mu}+\delta^I_{\;i}\delta^i_{\;\mu}\right],\quad \omega_{\mu}^{IJ}=
         h(t) \left(\delta^I_{\;0}\delta^J_{\;\mu}-\delta^I_{\;\mu}\delta^J_{\;0}\right).
\end{equation}
In this way, we obtain the symmetry-reduced action
\begin{equation}\label{Eq:Symmetry-reduced-Action}
    S=3V_0\int \de t\left[M_{\rm Pl}^2\left(a^2(\dot{h}+n \,h^2)+b^2(\dot{h}+n\,c\,h^2)\right)+\tilde{c}_1n\left(a^4+c\,b^4+a^2b^2(1+c)\right)\right]~.
\end{equation}
In order to avoid having a divergent integral, following standard practice we have introduced a fiducial cell by cutting off the integral over spatial leaves to a finite volume $V_0$ \cite{Ashtekar:2011ni}.

Varying the symmetry-reduced action \eqref{Eq:Symmetry-reduced-Action} with respect to the independent dynamical variables $a,\,b,\,c,\,n,\,h$, integrating by parts when necessary and disregarding boundary terms, we obtain the following set of equations of motion
\begin{subequations}\label{Eq:Eom-SRA}
\begin{equation}\label{Eq:Eom-SRA-a}
    M_{\rm Pl}^2(\dot{h}+n\, h^2)+\tilde{c}_1 n\left[2\,a^2+b^2(1+c)\right]=0~,
\end{equation}
\begin{equation}\label{Eq:Eom-SRA-b}
    M_{\rm Pl}^2(\dot{h}+n\,c\,h^2)+\tilde{c}_1 n\left[2\,c\,b^2+a^2(1+c)\right]=0~,
\end{equation}
\begin{equation} \label{Eq:Eom-SRA-c}
    h^2=-\frac{\tilde{c}_1}{M_{\rm Pl}^2}(a^2+b^2)~,
\end{equation}
\begin{equation}\label{Eq:Eom-SRA-n}
    M_{\rm Pl}^2(a^2+b^2c)h^2+\tilde{c}_1(a^2+b^2)(a^2+cb^2)=0~,
\end{equation}
\begin{equation}\label{Eq:Eom-SRA-h}
     h=\frac{a\dot{a}+b\dot{b}}{n(a^2+b^2c)}~.
\end{equation}
\end{subequations}
If equation \eqref{Eq:Eom-SRA-c} holds, then \eqref{Eq:Eom-SRA-n} becomes trivial (since $a^2+b^2c\neq0$ on Branch~I). On the other hand, using \eqref{Eq:Eom-SRA-c}, equations \eqref{Eq:Eom-SRA-a} and \eqref{Eq:Eom-SRA-b} can be rewritten in the same form as \eqref{Eq:Cosmological-eom-b} and \eqref{Eq:Cosmological-eom-c}. Hence, the system of equations \eqref{Eq:Eom-SRA} obtained from the symmetry-reduced action is completely equivalent to the system \eqref{Eq:Cosmological-eom} derived from the general action (which, as shown earlier, further simplifies to \eqref{Eq:Cosmological-eom-final}). Hence, we conclude that the {\it symmetric criticality} holds in this framework when restricted to Branch I. Therefore, in the following we will consider the symmetry-reduced action \eqref{Eq:Symmetry-reduced-Action} as the starting point to carry out a detailed analysis of the dynamical properties of the system at hand and gain a deeper understanding of its gauge symmetries~\eqref{Eq:Solution-NCG-branchI}.

\subsection{Canonical analysis}\label{Sec:CanonicalAnalysis}

Gauge freedom is more clearly displayed in the Hamiltonian formalism. For this reason, we will transform our singular Lagrangian system \eqref{Eq:Symmetry-reduced-Action} into Hamiltonian form following the Dirac-Bergmann algorithm \cite{dirac1964lectures,Henneaux:1992ig}.
First, we introduce the conjugate momenta, defined as $\pi_{\alpha}\coloneqq{\delta S}/{\delta \dot{\alpha}}$\,, where $\alpha$ denotes any of the dynamical variables. We obtain
\begin{equation}
    \pi_a=\pi_b=\pi_c=\pi_n=0~,\qquad \pi_h=3V_0 M_{\rm Pl}^2(a^2+b^2)~.
\end{equation}
These define the primary constraints
\begin{equation}
   \phi_1\coloneqq \pi_a=0~,\quad \phi_2\coloneqq\pi_b=0~,\quad \phi_3\coloneqq\pi_c=0~,\quad \phi_4\coloneqq\pi_n=0~,\quad \phi_5\coloneqq\pi_h-3V_0M_{\rm Pl}^2(a^2+b^2)=0~.
\end{equation}
The next step is to compute the canonical Hamiltonian, $H_C$\,, by performing a Legendre transformation of the Lagrangian, and expressing the result in terms of the phase space variables $\alpha$\,, $\pi_\alpha$\,. We obtain
\begin{equation}H_C\coloneqq\pi_\alpha\dot{\alpha}-L= -n\,V_0\,M_{\rm Pl}^2\,(a^2+b^2c)(3h^2+ \tilde{c}_1\,M_{\rm Pl}^{-4}\,V_0^{-1}\pi_h)~,
\end{equation}
From here we obtain the primary Hamiltonian $H_P$ as the sum of $H_C$ and all the primary constrains $\phi_{i}$ with Lagrange multipliers $\lambda_{i}$\,, that is, $H_P\coloneqq H_C+\lambda_i\phi_i$\,, with $i=1,\dots,5$\,. Next, we need to ensure that the primary constraints are preserved by time evolution. This is achieved by imposing the consistency conditions $\dot{\phi}_i=[\phi_i,H_P]=0$\,, where, along this section, $[\,,]$ denotes the Poisson bracket, defined as usual by $[F,G]=\frac{\partial F}{\partial \alpha}\frac{\partial G}{\partial \pi_\alpha}-\frac{\partial F}{\partial \pi_\alpha}\frac{\partial G}{\partial \alpha}$ \cite{Henneaux:1992ig}. These lead to one secondary constraint
\begin{equation}
    \phi_6\coloneqq h^2+\frac{\tilde{c}_1}{3}M_{\rm Pl}^{-4}V_0^{-1}\pi_h=0~,
\end{equation}
and two conditions over the Lagrange multipliers, $\lambda_5=0$\,, and $a\lambda_1+b\lambda_2=n\,h(a^2+b^2c)$\,. As the secondary constraint itself satisfies the consistency condition $\dot{\phi}_6=0$\,, there are neither tertiary constraints nor further restrictions on the Lagrange multipliers and the process naturally stops. 

At this stage, it is convenient to introduce the weak equality symbol ``$\approx$'' to emphasize that the constraints $\phi_j$\,, with $j=1,\dots,6$\,, are restricted to be zero only on the constraint hypersurface, but do not identically vanish throughout phase space. Hence, given two functions $F,\,G$ we will write $F\approx G$ to denote that they coincide on the submanifold defined by the constraints $\phi_j\approx 0$ \cite{Henneaux:1992ig}.

The ensuing point is to obtain the total Hamiltonian, $H_T$\,, which is defined from the primary Hamiltonian, $H_P$\,, by inserting the general solution for the Lagrange multipliers:
\begin{equation}
    H_T=\frac{n(a^2+c b^2)}{a \,M_{\rm Pl}^2}\left(M_{\rm Pl}^2\, h \,\pi_a-\tilde{c}_1 a \,\pi_h-3\,a\, M_{\rm Pl}^4V_0\, h^2\right)+\frac{\lambda_2}{a}(a\,\pi_b-b\,\pi_a)+\lambda_3\pi_c+\lambda_4\pi_n~.
\end{equation}
In terms of the total Hamiltonian, physical phase-space orbits of the system are defined by $\dot{F}\approx [F,H_T]$\,, with initial data that satisfy $\phi_j\approx 0$\,. These equations contain three arbitrary functions $\lambda_2$, $\lambda_3$, and $\lambda_4$, and are equivalent, by construction, to the equations of motion \eqref{Eq:Eom-SRA}. The presence of these three functions indicates that there is more than one set of values of the canonical variables representing the same physical state.

In the process of identifying the gauge symmetries of the system, the crucial aspect lies in the classification of constraints into first and second class. First-class constraints are those that have weakly vanishing Poisson brackets with all the constraints. Second-class constraints are those that are not first-class, and thus can be grouped into pairs of constraints with non-vanishing Poisson brackets. The matrix $C_{jk}\coloneqq[\phi_j,\phi_k]$\,, with $j,k=1,...,6$ is given by 
\begin{equation}
    C=\begin{pmatrix}
0 & 0 & 0 & 0 & 6\sqrt{3}M_{\rm Pl}^3V_0\, a & 0\\
0 & 0 & 0 & 0 & 6\sqrt{3}M_{\rm Pl}^3V_0 \,b & 0\\
0 & 0 & 0 & 0 & 0 & 0 \\
0 & 0 & 0 & 0 & 0 & 0 \\
-6\sqrt{3}M_{\rm Pl}^3V_0\, a & -6\sqrt{3}M_{\rm Pl}^3V_0\, b & 0 & 0 & 0 & -2M_{\rm Pl}^{-1}h/\sqrt{3}\\
0 & 0 & 0 & 0 & 2 M_{\rm Pl}^{-1}h/\sqrt{3} & 0 \\
    \end{pmatrix}~,
\end{equation}
which is non-zero on the constraint surface. Applying Theorem 1.3 in \cite{Henneaux:1992ig}, we obtain 4 fist-class constraints
\begin{equation}\label{Eq:First-class}
    \Phi_1\coloneqq a\pi_b-b\pi_a\approx 0~,\quad \Phi_2\coloneqq \pi_c\approx 0~,\quad\Phi_3\coloneqq \pi_n\approx 0~,\quad \Phi_4\coloneqq 3\,b \,h^2\,V_0 \,M_{\rm Pl}^4-h\,M_{\rm Pl}^2\pi_b+ \tilde{c}_1 b\,\pi_h\approx 0~,
\end{equation}
as well as 2 second-class constraints
\begin{equation}\label{Eq:Second-class}
    \Phi_5\coloneqq 3^3(a\pi_a+b\pi_b)M_{\rm Pl}^8\, V_0+M_{\rm Pl}^2\,h^3+ \tilde{c}_1(a^2+b^2)\,h\approx 0~,\quad \Phi_6\coloneqq 3(a^2+b^2)M_{\rm Pl}^2V_0-\pi_h\approx 0~.
\end{equation}
From these, $\Phi_1,\,\Phi_2,\,\Phi_3$ and $\Phi_6$ are primary constraints since they are constructed entirely from primary constraints, whereas $\Phi_4$ and $\Phi_5$ are secondary constraints.

Second-class constraints can be used to eliminate entirely some canonical variables from the description. Solving the constraints \eqref{Eq:Second-class} as strong equations (i.e., restricting to the hypersurfaces they define in phase space), we can eliminate, for instance, $\pi_b$ and $\pi_h$\,. After imposing second-class constraints strongly, Poisson brackets no longer preserve the constraints under time evolution. At this point, the Poisson bracket must be replaced by the Dirac bracket \cite{Henneaux:1992ig}, defined as $[F,G]_*\coloneqq [F,G]-[F,\chi_\alpha]\mathcal{M}^{\alpha\beta}[\chi_\beta,G]$\,, where $\chi_\alpha$ is the set of second class constraints and $\mathcal{M}^{\alpha\beta}$ is the inverse of the matrix of the Poisson brackets of all the second-class constraints. In reduced phase space, the first-class constraints read
\begin{subequations}
\begin{align}
     \Phi_1 & =(a^2+b^2) (\tilde{c}_1a\,h+3^3V_0\,M_{\rm Pl}^8\,\pi_a)+a\,M_{\rm Pl}^2\,h^3\approx 0~,\\
    \Phi_2 & =\pi_c\approx 0~,\\
    \Phi_3 & =\pi_n\approx 0~,\\
     \Phi_4 & =3^3a\,h\,V_0\,M_{\rm Pl}^8\,\pi_a+[ \tilde{c}_1(a^2+b^2)+M_{\rm Pl}^2\,h^2](h^2+3^4b^2\,V_0^2\,M_{\rm Pl}^8)\approx 0~.
\end{align} 
\end{subequations}

Gauge transformations at a given time are generated by the set of four first-class constraints listed above.\footnote{First-class constraints---primary as well as secondary---generate gauge transformations at a given time, understood as transformations that map a set of initial conditions
into a gauge-equivalent set. These should not be confused with the usual notion of gauge transformations as maps from solutions of the equations of motion into other solutions for the entire trajectory
\cite{Pons:2004pp}.}
In other words, an infinitesimal (point) gauge transformation is generated by a linear combination $\bar{\Phi}=\epsilon_m\Phi_m$\,, with $m=1,\dots,4$\,. The canonical variables transform as
\begin{subequations}
\begin{align}
    \delta_\varepsilon a & =\varepsilon_m[a,\Phi_m]_*=-b\varepsilon_1~,\\
    \delta_\varepsilon b & =\varepsilon_m[b,\Phi_m]_*=a\varepsilon_1-3\,h\, M_{\rm Pl}^2\,\varepsilon_4~,\\
    \delta_\varepsilon h & =\varepsilon_m[h,\Phi_m]_*=3\,\tilde{c}_1b\,\varepsilon_4~,\\
    \delta_\varepsilon c & =\varepsilon_m[c,\Phi_m]_*=\varepsilon_2~,\\
    \delta_\varepsilon n & =\varepsilon_m[n,\Phi_m]_*=\varepsilon_3~,
\end{align}
\end{subequations}
where the arbitrary gauge parameters $\varepsilon_m$ are functions of time. Thus, in the reduced phase space, the symmetries associated with the generators $\Phi_m$ correspond to: a rotation in the $(a,b)$ plane generated by $\Phi_1$, reparametrizations of the relative lapse $c$ as generated by $\Phi_2$, global time reparametrizations generated by $\Phi_3$,  and rescalings of $h$ generated by $\Phi_4$.

Once all the gauge transformations have been identified, counting the number of physical degrees of freedom becomes straightforward. The key point is that the canonical gauge-fixing procedure requires the same number of new constraints as the number of first-class constraints in the theory (so that the gauge-fixing constraints form pairs of second-class constraints with the corresponding first-class constraints). For a satisfactory set of gauge conditions, one is left with second-class constraints only, and therefore the total Hamiltonian is uniquely determined \cite{Gracia:1988xp}. Consequently, the initial ten-dimensional phase space undergoes a reduction by two due to the presence of the second-class constraints $\Phi_5$ and $\Phi_6$ that restrict the initial data, in addition to twice the number of first-class constraints, i.e., $2 \times 4 = 8$\,. Hence, we are led to the conclusion that the system has exactly zero physical degrees of freedom. Thus, the dynamics for this system in the pure gravity case are pure gauge. 

On a technical point, we note that we have not enlarged the dynamics by adding to the total Hamiltonian the secondary first-class constraint to build the extended Hamiltonian.
This is because in our case, the Dirac conjecture does not hold, in the sense that the time evolution derived from the original Lagrangian does not explicitly exhibit all the transformations that do not change the physical state of the system at a given time \cite{Henneaux:1992ig}.
Therefore, the extended Hamiltonian dynamics formalism fails to provide a correct description of the dynamics.

A direct comparison with GR further clarifies the peculiarity of the symmetry structure found here. In pure gravity GR, i.e.~in the absence of matter fields, the dynamics of homogeneous and isotropic FLRW spacetimes also possesses no dynamical degrees of freedom. Specifically, after imposing spatial symmetry, the description of the system becomes pure gauge, as highlighted by the Hamiltonian analysis, which yields a vanishing number of physical degrees of freedom. However, there is an important difference: while in GR, after fixing the time gauge, we are left with a uniquely determined functional form for the scale factor as a function of time (through the Friedmann equation), in the present model, the two independent scale factors remain completely arbitrary even after fixing the time gauge. This is a consequence of the extra gauge freedom in the model at hand, corresponding to the three extra first-class constraints, and is manifest in the solution \eqref{Eq:Solution-NCG-branchI}.

\section{Discussion}\label{Sec:Discussion}
In this work, we analysed the dynamics of the cosmological sector of a twist-deformation of general relativity in pure gravity (i.e.,~no matter sources are present), in the commutative limit, which is relevant for the dynamics on scales much larger than the noncommutativity scale. The theory was proposed in Ref.~\cite{deCesare:2018cjr}, as an extension of Ref.~\cite{Aschieri:2009ky} by means of additional interaction terms compatible with the underlying ${\rm GL}(2,\mathbb{C})$ symmetry. In the commutative limit, one obtains an effective bigravity theory \ref{Eq:NCG-Action} with two tetrads, $e^I$, $\tilde{e}^I$, and a single spin connection $\omega^{IJ}$, plus an additional one-form $A$ playing the role of a Lagrange multiplier implementing the constraint $\de(e^I\wedge\tilde{e}_I)=0$\,.
We consider a general ansatz corresponding to a pair of spatially flat FLRW geometries with two independent scale factors, in a general time gauge, and with arbitrary relative lapse between their respective conformal times.
Cosmological solutions split into two branches, depending on whether the two-form $e^I\wedge e^J+\tilde{e}^I\wedge\tilde{e}^J$ has a non-zero time-space component (Branch I) or is purely spatial (Branch II). We analysed the two branches separately.

On Branch I, the dynamics display rich structures. The spin connection depends on a scalar function $h(t)$, which plays the role of an effective Hubble rate and satisfies an effective Friedmann equation~\eqref{Eq:Cosmological-eom-final}. The solution to the equations of motion in a general time gauge \eqref{Eq:Solution-NCG-branchI} depends upon three arbitrary functions of time. We investigate further the origin of this phenomenon in Section~\ref{Sec:SymRed}, where it is shown that it originates from an enhanced gauge symmetry compared to general relativity. Specifically, after showing that the symmetric criticality principle holds on this branch in Section~\ref{Sec:SymRed_sub1}, we performed a rigorous Hamiltonian analysis for the symmetry-reduced theory in the cosmological sector in Section~\ref{Sec:CanonicalAnalysis}. We show that the Hamiltonian description of the dynamics features four first-class constraints and two second-class constraints. By solving the second-class constraints as strong equations and computing the Dirac bracket, we move to the reduced phase space. This allows us to show that the four first-class constraints are associated with gauge symmetries corresponding to global time reparametrizations, reparametrizations of the relative lapse between conformal times, rescalings of the Hubble rate, and rotations in the $(a,b)$ plane spanned by the two scale factors.

On Branch II, the dynamics of the scale factors associated with the two tetrads correspond to the motion of a point-particle on a circle with constant radius, with arbitrary velocity. The curvature of the spin connection is constant and purely spatial. However, the evolution of both scale factors remains arbitrary, and the torsion is nonzero. Only in the special case where the Hubble rate is tuned to a particular value, such that its square coincides with the field strength, torsion vanishes and one recovers a de Sitter geometry. This branch is of limited interest for the purpose of this study.

In the Appendix~\ref{Sec:ComparisonAS} we performed a detailed comparison between the cosmological dynamics in the model at hand and the Alexandrov-Speziale model, with action \eqref{Eq:Speziale-Action}. The interaction terms between tetrads in the action~\eqref{Eq:ActionNCG} correspond to a special case of the bigravity interactions considered in \eqref{Eq:Speziale-Action}, such that there is a single free coupling determining both the quadratic interaction and quartic self-interaction terms. However, in \eqref{Eq:ActionNCG} there is an extra term implementing the constraint $\de(e^I\wedge\tilde{e}_I)=0$\,, which is absent from \eqref{Eq:Speziale-Action}.
Our analysis shows that the Alexandrov-Speziale does not share the enhanced symmetry of \eqref{Eq:ActionNCG}, which originates precisely from the particular tuning of the interaction couplings. In this theory too, solutions can be divided into two branches. In both cases, the general solution describes a pair of de~Sitter universes with an effective cosmological constant determined by the coupling constants.

Future work should aim at a deeper understanding of the physical degrees of freedom in the model at hand, going beyond the FLRW background and the pure gravity case here considered.
To this end, it is necessary to extend the analysis presented here to the general case, to identify the physical degrees of freedom and determine their dynamical properties
beyond highly symmetric geometries and in the presence of matter fields. This is a necessary step for the purpose of building realistic cosmological models. Moreover, a comparison of perturbative and nonperturbative analyses of the propagating degrees of freedom may reveal whether the extra degrees of freedom may be `hidden' around certain backgrounds, as in the model of Ref.~\cite{Alexandrov:2019dxm}.

We note that the peculiar structure of the interaction terms between tetrads in the action \eqref{Eq:NCG-Action} is the same as in the partially massless candidate theory proposed in Ref.~\cite{Hassan:2012gz} in the framework of Hassan-Rosen bigravity. However, the remaining terms of the gravitational action are different in the two theories, as noted above, and the action \eqref{Eq:NCG-Action} has an additional gauge symmetry at the full non-linear level.
Therefore, it would be interesting to investigate whether the no-go theorems \cite{Joung:2014aba,Apolo:2016vkn}, which prevent the extension of the partially massless gauge symmetry to the full nonlinear level, could be circumvented in the present framework.
This is left for future work.

Lastly, the emergence of bigravity theories in the twist-deformation approach to noncommutative geometry appears to be rather general when a first-order formulation of the dynamics is adopted, which may offer guidance for building fundamentally motivated alternative theories of gravity. In addition, this prompts further investigations to extend the present analysis to other gravity theories considered in this framework, such as e.g.~the noncommutative deformation of the MacDowell-Mansouri model considered in Ref.~\cite{Aschieri:2009ky} and the ${\rm SO}(2,3)_{\star}$ theory of Ref.~\cite{DimitrijevicCiric:2016qio}, to shed light on their dynamics in the commutative limit.


\appendix

\section{Comparison with the Alexandrov-Speziale theory}\label{Sec:ComparisonAS}

In this appendix we aim to continue the background analysis of the theory presented in Ref.~\cite{Alexandrov:2019dxm}, with action \eqref{Eq:Speziale-Action}, and draw a closer comparison with the theory at hand \eqref{Eq:NCG-Action}. Specifically, we will focus on cosmological solutions, which for \eqref{Eq:NCG-Action} have been analyzed in Section~\ref{Sec:NGC-cosmo}. 
We start by observing that the variation of the action \eqref{Eq:Speziale-Action} with respect to the spin connection $\omega^{IJ}$ gives the same connection equation \eqref{Eq:Eom-a} as in the model \eqref{Eq:NCG-Action}. The remaining equations of motion derived from this action are
\begin{subequations}\label{Eq:Speziale-Eom}
    \begin{equation} \label{Eq:Speziale-Eom-a}
        \epsilon_{IJKL}\left(e^J\wedge F^{KL}-\frac{\beta_3}{3}\tilde{e}^J\wedge \tilde{e}^K\wedge \tilde{e}^L-\beta_2e^J\wedge \tilde{e}^K\wedge \tilde{e}^L-\beta_1 e^J\wedge e^K\wedge \tilde{e}^L-\frac{\beta_0}{3}e^J\wedge e^K\wedge e^L\right)=0~,
    \end{equation}
     \begin{equation} \label{Eq:Speziale-Eom-b}
        \epsilon_{IJKL}\left(\tilde{e}^J\wedge F^{KL}-\frac{\beta_4}{3}\tilde{e}^J\wedge \tilde{e}^K\wedge \tilde{e}^L-\beta_3e^J\wedge \tilde{e}^K\wedge \tilde{e}^L-\beta_2 e^J\wedge e^K\wedge \tilde{e}^L-\frac{\beta_1}{3}e^J\wedge e^K\wedge e^L\right)=0~.
    \end{equation}
\end{subequations}
In Ref.~\cite{Alexandrov:2019dxm} these equations are solved for a general class conformally-equivalent tetrad backgrounds, such that $e^{I}=\Omega\, \check{e}^I$, $\tilde{e}^{I}=\tilde{\Omega}\, \check{e}^I$ (with reference tetrad $\check{e}^I$ and conformal factors $\Omega$\,, $\tilde{\Omega}$). Here, we are going to consider a different ansatz, which allows for more generality in the cosmological case 
\begin{equation}\label{Eq:Tetrads-spe}
    e^I_{\;\mu}=a(t)\left(n(t)\delta^J_{\;0}\delta^0_{\;\mu}+\delta^J_{\;i}\delta^i_{\;\mu}\right),\qquad \tilde{e}^I_{\;\mu}=b(t)\left(n(t)c(t)\delta^I_{\;0}\delta^0_{\;\mu}+\delta^I_{\;i}\delta^i_{\;\mu}\right)~.
\end{equation}
We note that, in general, the two tetrads in \eqref{Eq:Tetrads-spe} are not conformally equivalent in general, unless the relative lapse is $c(t)=1$\,.

Since the connection equation is exactly the same as in the NCG model~\eqref{Eq:ActionNCG}, which we analyzed in Section~\ref{Sec:NGC-cosmo}, and the ansatz \eqref{Eq:Tetrads-spe} is also the same as \eqref{Eq:FLRW-ansatz}, in this case too we can distinguish between two branches of solutions, depending on the value of $a^2+c b^2$.
 These are analyzed in the remainder of this Appendix. 

\subsection{Branch I}
Solutions on this branch are such that $a^2+ b^2c\neq 0$\,.
This case displays several analogies with the corresponding branch of Section~\ref{Subsec:BranchII-NCG}. Here too, the solution to the spin connection can be expressed as $\omega_{\mu}^{IJ}=h(t) \left(\delta^I_{\;0}\delta^J_{\;\mu}-\delta^I_{\;\mu}\delta^J_{\;0}\right)$\,, where once again we use the shorthand notation $h$
\begin{equation}
    h\coloneqq\frac{a\dot{a}+b\dot{b}}{n(a^2+cb^2)}~.
\end{equation}
The remaining equations of motion read
\begin{subequations}\label{Eq:Speziale-FLRW}
    \begin{equation}\label{Eq:Speziale-FLRW-a}
        3 ah^2-\beta_3 b^3-3\beta_2 a b^2-3\beta_1 a^2b-\beta_0 a^3=0~,
    \end{equation}
    \begin{equation}\label{Eq:Speziale-FLRW-b}
        2a\dot{h}+ an h^2-n\left[\beta_3 c b^3-\beta_2 a b^2(1+2c)-\beta_1 a^2b(2+c)-\beta_0 a^3\right]=0~,
    \end{equation}
    \begin{equation}\label{Eq:Speziale-FLRW-c}
        3 b h^2-\beta_4 b^3-3\beta_3 a b^2-3\beta_2 a^2b-\beta_1 a^3=0~,
    \end{equation}
    \begin{equation}\label{Eq:Speziale-FLRW-d}
        2b\dot{h}+ bcn h^2-n\left[\beta_4 c b^3-\beta_3 a b^2(1+2c)-\beta_2 a^2b(2+c)-\beta_1 a^3\right]=0~.
    \end{equation}
\end{subequations}
Thus, for the three unknowns $a$, $b$, $c$ we have, in principle, four equations. In order to see whether the system is compatible or not, let us use again the notation $y\coloneqq b/a$ and rewrite \eqref{Eq:Speziale-FLRW} as 
\begin{subequations}\label{Eq:Speziale-FLRW-y}
\begin{equation}\label{Eq:Speziale-FLRW-y-a}
        \frac{3h^2}{a^2}-\beta_3 y^3-3\beta_2 y^2-3\beta_1 y-\beta_0=0~,
    \end{equation}
    \begin{equation}\label{Eq:Speziale-FLRW-y-b}
        \frac{3h^2}{a^2}-\beta_4 y^2-3\beta_3 y-3\beta_2-\beta_1 y^{-1}=0~,
    \end{equation}
\begin{equation}\label{Eq:Speziale-FLRW-y-c}
         \frac{2\dot{h}}{a^2 n}=-\frac{h^2}{a^2}+\beta_3 c y^3+\beta_2 y^2(1+2c)+\beta_1 y(2+c)+\beta_0~,
    \end{equation}
    \begin{equation}\label{Eq:Speziale-FLRW-y-d}
        \frac{(c-1)h^2}{a^2}+ \beta_3cy^3+\left[(1+2c)\beta_2-c\beta_4\right]y^2+\left[(2+c)\beta_1-(1+2c)\beta_3\right]y+\beta_0-(2+c)\beta_2-\beta_1y^{-1}=0~.
    \end{equation}
\end{subequations}
Then, from \eqref{Eq:Speziale-FLRW-y-a} and \eqref{Eq:Speziale-FLRW-y-b} 
\begin{equation}\label{Eq:y-polynomial}
    \frac{\beta_3}{3}y^{4}+\left(\beta_2-\frac{\beta_4}{3}\right)y^{3}+\left(\beta_1-\beta_3\right)y^{2}+\left(\frac{\beta_0}{3}-\beta_2\right)y-\frac{\beta_1}{3}=0~.
\end{equation}
Interestingly, this is the same equation obtained in Hassan-Rosen bimetric gravity in the absence of matter sources \cite{vonStrauss:2011mq}.
Moreover, for the particular choice of the couplings $\beta_n$ that reduces \eqref{Eq:Speziale-Action} to \eqref{Eq:NCG-Action}, Eq.~\eqref{Eq:y-polynomial} becomes an identity. In this case, the solution to the system \eqref{Eq:Speziale-FLRW-y} has been analyzed in previous sections and is given by \eqref{Eq:Solution-NCG-branchI}. For any other choices of the couplings $\beta_n$\,, the polynomial equation \eqref{Eq:y-polynomial} admits non-trivial solutions $y=y_*$ with $y_*={\rm constant}$\,, which in turn imply $b=y_*a$\,. 

The case $c=1$\,, corresponding to conformally equivalent tetrads, is special and requires a separate discussion. In this case, \eqref{Eq:y-polynomial} and \eqref{Eq:Speziale-FLRW-y-d} become identical, and the remaining independent equations of the system \eqref{Eq:Speziale-FLRW-y} reduce to
\begin{subequations}
    \begin{align}
         \frac{2\dot{h}}{a^2n}+\frac{h^2}{a^2}&=\beta_3  y_*^3+3\beta_2 y_*^2+3\beta_1 y_*+\beta_0~,\\
        \frac{3h^2}{a^2}&=\beta_3 y_*^3+3\beta_2 y_*^2+3\beta_1 y_*+\beta_0~,
    \end{align}
\end{subequations}
with $y_*$ given by \eqref{Eq:y-polynomial}. Since in this case $h=\dot{a}/(n\,a)$\,, these are equivalent to the standard Friedmann equations for a spatially flat universe (with scale factor $a$ and lapse $n$) and an effective cosmological constant $\Lambda_{\rm eff}=\beta_3 y_*^3+3\beta_2 y_*^2+3\beta_1 y_*+\beta_0$\,.
Hence, we only have one independent equation, 
\begin{equation}\label{Eq:Speziale-c1}
        \frac{3h^2}{a^2}=\beta_3 y_*^3+3\beta_2 y_*^2+3\beta_1 y_*+\beta_0~,
    \end{equation}
whose solution is a de Sitter Universe for both metrics. This is consistent with the results obtained earlier in \cite{Alexandrov:2019dxm} for general conformally related tetrads (which includes our cosmological ansatz \eqref{Eq:FLRW-ansatz} when $c=1$).

On the other hand, in the  more general case $c\neq 1$\,---not included in the analysis of Ref.~\cite{Alexandrov:2019dxm}, which focuses on conformally equivalent tetrads---we can combine \eqref{Eq:Speziale-FLRW-y-a} and \eqref{Eq:Speziale-FLRW-y-d} to derive an algebraic equation for $c$\,:
\begin{multline}\label{Eq:c-polynomial}
      \beta_3cy_*^3+\left[(1+2c)\beta_2-c\beta_4\right]y_*^2+\left[(2+c)\beta_1-(1+2c)\beta_3\right]y_*+\beta_0-(2+c)\beta_2-\beta_1y_*^{-1}\\
      =\frac{1-c}{3}\left[\beta_3 y_*^3+3\beta_2 y_*^2+3\beta_1 y_*+\beta_0\right]~,
\end{multline}
which gives $c=c_*$ with $c_*={\rm constant}$\,. Using \eqref{Eq:y-polynomial}, this equation can be rewritten as
\begin{equation}\label{Eq:c-polynomial-2}
    (c_*-1)\left[\beta_1+3(\beta_1-\beta_3)y_*^2+2(3\beta_2-\beta_4)y_*^3+3\beta_3y_*^4\right]=0~.
\end{equation}
Since we are assuming $c_*\neq 1$\,, the combination of terms enclosed by square brackets must vanish. This constitutes a constraint on the parameters $\beta_n$\,.
Hence, in this case, the solution is
\begin{subequations}\label{Eq:Speziale-FLRW-solution}
\begin{align}
        \frac{3h^2}{a^2}&=\beta_3 y_*^3+3\beta_2 y_*^2+3\beta_1 y_*+\beta_0~,\label{Eq:Speziale-FLRW-solution-a}\\
         \frac{2\dot{h}}{a^2 n}+\frac{h^2}{a^2}&=\beta_3 c_* y_*^3+\beta_2 y_*^2(1+2c_*)+\beta_1 y_*(2+c_*)+\beta_0~,\label{Eq:Speziale-FLRW-solution-b}
\end{align}
\end{subequations}
with the constants $y_*$ and $c_*$ given by Eqs.~\eqref{Eq:y-polynomial} and \eqref{Eq:c-polynomial}, respectively. Also in this case, we have only one independent equation, as can be easily checked by differentiating \eqref{Eq:Speziale-FLRW-solution-a} with respect to time, which gives
\begin{equation}
    \dot{h}=\frac{h\dot{a}}{a}=\frac{1+c_*y_*^2}{1+y_*^2}n\,h^2~,
\end{equation}
and therefore
\begin{equation}
    \frac{2\dot{h}}{a^2n}+\frac{h^2}{a^2}=\left(\frac{2(1+c_*y_*^2)}{1+y_*^2}+1\right)\frac{h^2}{a^2}=\frac{1}{3}\left(\frac{2(1+c_*y_*^2)}{1+y_*^2}+1\right)(\beta_3 y_*^3+3\beta_2 y_*^2+3\beta_1 y_*+\beta_0)~.
\end{equation}
This is equivalent to \eqref{Eq:Speziale-FLRW-solution-b} when both constraints \eqref{Eq:y-polynomial} and \eqref{Eq:c-polynomial} hold. Hence, also for $c_*\neq 1$ we obtain a de~Sitter Universe, this time with effective cosmological constant $\tilde{\Lambda}_{\rm eff}=\frac{(1+c_*y_*^2)^2}{(1+y_*^2)^2}\left(\beta_3 y_*^3+3\beta_2 y_*^2+3\beta_1 y_*+\beta_0\right)$\,.

\subsection{Branch II}
For $a^2+b^2c=0$\,, the solution to the connection equation is given by \eqref{Eq:Connection-branchI-sol}. On the other hand, specializing the field equations \eqref{Eq:Speziale-Eom} to our ansatz for the tetrads \eqref{Eq:Tetrads-spe}, we obtain the following set of equations for the remaining independent components of the spin connection
\begin{subequations}
\begin{equation}\label{Eq:BranchI-spe-a}
  \omega^0_{11}\omega^0_{22}+\omega^0_{11}\omega^0_{33}+\omega^0_{22}\omega^0_{33}-(\omega^0_{12})^2-(\omega^0_{13})^2-(\omega^0_{23})^2-b\left(\beta_0\frac{a^2}{b}+3\beta_1 a+3\beta_2 b+\beta_3\frac{b^2}{a}\right)=0~,
\end{equation} 
\begin{equation}\label{Eq:BranchI-spe-b}
    \omega^0_{ii}\omega^0_{jj}-(\omega^0_{ij})^2+\frac{1}{n}\left(\Dot{\omega}^0_{ii}+\Dot{\omega}^0_{jj}\right)-\left(2\beta_1-\beta_3\right)a\,b-\left(\beta_0-2\beta_2\right)a^2-\beta_2b^2+\beta_1\frac{a^3}{b}=0~,
\end{equation}
\begin{equation}\label{Eq:BranchI-spe-c}
    \omega^0_{ii}\omega^0_{jk}-\omega^0_{ij}\omega^0_{ik}+\frac{1}{n}\Dot{\omega}^0_{jk}=0~,
\end{equation}
\begin{equation}\label{Eq:BranchI-spe-d}
  \omega^0_{11}\omega^0_{22}+\omega^0_{11}\omega^0_{33}+\omega^0_{22}\omega^0_{33}-(\omega^0_{12})^2-(\omega^0_{13})^2-(\omega^0_{23})^2-a\left(\beta_1\frac{a^2}{b}+3\beta_2 a+3\beta_3 b+\beta_4\frac{b^2}{a}\right)=0~,
\end{equation} 
\begin{equation}\label{Eq:BranchI-spe-e}
    \omega^0_{ii}\omega^0_{jj}-(\omega^0_{ij})^2-\frac{b^2}{a^2 n}\left(\Dot{\omega}^0_{ii}+\Dot{\omega}^0_{jj}\right)+\left(\beta_1-2\beta_3\right)a\,b+\left(2\beta_2-\beta_4\right)b^2-\beta_2a^2+\beta_3\frac{b^3}{a}=0~,
\end{equation}
\begin{equation}\label{Eq:BranchI-spe-f}
    \omega^0_{ii}\omega^0_{jk}-\omega^0_{ij}\omega^0_{ik}-\frac{b^2}{a^2n}\Dot{\omega}^0_{jk}=0~.
\end{equation}
\end{subequations}
From \eqref{Eq:BranchI-spe-c} and \eqref{Eq:BranchI-spe-f} it follows that necessarily $\omega^0_{jk}$ must be constant. Combining \eqref{Eq:BranchI-spe-a} and \eqref{Eq:BranchI-spe-d}, we derive the following polynomial equation for $y\coloneqq b/a$\,, 
\begin{equation}\label{Eq:BranchI-spe-y}
    -\beta_1+(\beta_0-3\beta_2)y+3(\beta_1-\beta_3)y^2+(3\beta_2-\beta_4)y^3+\beta_3y^4=0~,
\end{equation}
The special case where the $\beta_n$ are such that \eqref{Eq:BranchI-spe-y} is an identity corresponds to the NCG model (with $\beta_1=\beta_3=0$\,, $\beta_0=\beta_4=3\beta_2$) and has been analyzed in Section~\ref{Subsec:BranchI-NCG}. For general values of the $\beta_n$\,, Eq.~\eqref{Eq:BranchI-spe-y} is an algebraic equation for $y$\,, whose solution gives $b/a={\rm constant}$.
Along with the constraint $a^2+b^2=\mathcal{C}$\,, this implies that both $a$ and $b$ must be constant. From \eqref{Eq:BranchI-spe-a} and \eqref{Eq:BranchI-spe-d} one then obtains that $\omega^0_{ii}$ must be constant too. Hence, we conclude that all the components of the spin connection must be constant. Finally, solving \eqref{Eq:BranchI-spe-b} along with \eqref{Eq:BranchI-spe-c} we obtain that, as in the NCG model, $\omega^0_{11}=\omega^0_{22}=\omega^0_{33}$ and $\omega^0_{12}=\omega^0_{13}=\omega^0_{23}=0$\,, which leads to a purely spatial field strength.

\bibliographystyle{bib-style}
\bibliography{references_NCG}

\end{document}